\documentclass[12pt,a4paper]{article}
\usepackage{ifthen}
\usepackage{booktabs}
\usepackage{overpic}
\usepackage{relsize}
\usepackage{rotating}
\newboolean{pdflatex}
\setboolean{pdflatex}{true}

\newboolean{articletitles}
\setboolean{articletitles}{true}

\newboolean{uprightparticles}
\setboolean{uprightparticles}{false}

\newboolean{inbibliography}
\setboolean{inbibliography}{false}

\usepackage{microtype}
\usepackage{lineno}  
\usepackage{xspace} 
\usepackage{caption}

\usepackage{graphicx}  
\usepackage{color}
\usepackage{colortbl}

\usepackage{amsmath} 
\usepackage{amssymb}
\usepackage{amsfonts}
\usepackage{upgreek} 

\newcommand*\patchAmsMathEnvironmentForLineno[1]{%
\expandafter\let\csname old#1\expandafter\endcsname\csname #1\endcsname
\expandafter\let\csname oldend#1\expandafter\endcsname\csname
end#1\endcsname
 \renewenvironment{#1}%
   {\linenomath\csname old#1\endcsname}%
   {\csname oldend#1\endcsname\endlinenomath}%
}
\newcommand*\patchBothAmsMathEnvironmentsForLineno[1]{%
  \patchAmsMathEnvironmentForLineno{#1}%
  \patchAmsMathEnvironmentForLineno{#1*}%
}
\AtBeginDocument{%
\patchBothAmsMathEnvironmentsForLineno{equation}%
\patchBothAmsMathEnvironmentsForLineno{align}%
\patchBothAmsMathEnvironmentsForLineno{flalign}%
\patchBothAmsMathEnvironmentsForLineno{alignat}%
\patchBothAmsMathEnvironmentsForLineno{gather}%
\patchBothAmsMathEnvironmentsForLineno{multline}%
\patchBothAmsMathEnvironmentsForLineno{eqnarray}%
}

\usepackage{hyperref}    
\usepackage[all]{hypcap} 

 \mathchardef\PLambda="7103
\def\pt         {\mbox{$p_{\rm T}$}\xspace}
\newcommand{\tev}{\ifthenelse{\boolean{inbibliography}}{\ensuremath{~T\kern -0.05em eV}\xspace}{\ensuremath{\mathrm{\,Te\kern -0.1em V}}}\xspace}
\newcommand{\gev}{\ensuremath{\mathrm{\,Ge\kern -0.1em V}}\xspace}
\newcommand{\mev}{\ensuremath{\mathrm{\,Me\kern -0.1em V}}\xspace}
\newcommand{\kev}{\ensuremath{\mathrm{\,ke\kern -0.1em V}}\xspace}

\def\fb   {\ensuremath{\mbox{\,fb}}\xspace}
\def\ps   {\ensuremath{\mbox{\,ps}}\xspace}
\def\invfb   {\ensuremath{\mbox{\,fb}^{-1}}\xspace}
\def\lhcb {\mbox{LHCb}\xspace}
\def\Pb      {\ensuremath{b}\xspace}
 \def\Pc      {\ensuremath{c}\xspace}
\def\bquark    {{\ensuremath{\Pb}}\xspace}
\def\cquark    {{\ensuremath{\Pc}}\xspace}

\def\Dbar    {{\kern 0.2em\overline{\kern -0.2em D}{}}\xspace}

\def\Dzb     {{\ensuremath{\Dbar{}^0}}\xspace}

\usepackage{cite} 
\usepackage{mciteplus}

\usepackage{longtable}
\usepackage{chngpage}

\def \smdecay {\ensuremath{B^0 \!\to K^{*0}\mu^+\mu^-}\xspace}
\def \sigdecay {\ensuremath{B^0 \!\to K^{*0}\chi}\xspace}
\def \mmm {\ensuremath{m(\mu^+\mu^-)}\xspace}
\def \tmm {\ensuremath{\tau(\mu^+\mu^-)}\xspace}
\def \mx {\ensuremath{m(\chi)}\xspace}
\def \tx {\ensuremath{\tau(\chi)}\xspace}
\def\jpsi     {{\ensuremath{{J\mskip -3mu/\mskip -2mu\psi\mskip 2mu}}}\xspace}
\def \mmmsq {\ensuremath{m^2(\mu^+\mu^-)}\xspace}

\begin{document}

\renewcommand{\thefootnote}{\fnsymbol{footnote}}
\setcounter{footnote}{1}

\begin{titlepage}
\pagenumbering{roman}

\vspace*{-1.5cm}
\centerline{\large EUROPEAN ORGANIZATION FOR NUCLEAR RESEARCH (CERN)}
\vspace*{1.5cm}
\hspace*{-0.5cm}
\begin{tabular*}{\linewidth}{lc@{\extracolsep{\fill}}r}
\ifthenelse{\boolean{pdflatex}}
{\vspace*{-2.7cm}\mbox{\!\!\!\includegraphics[width=.14\textwidth]{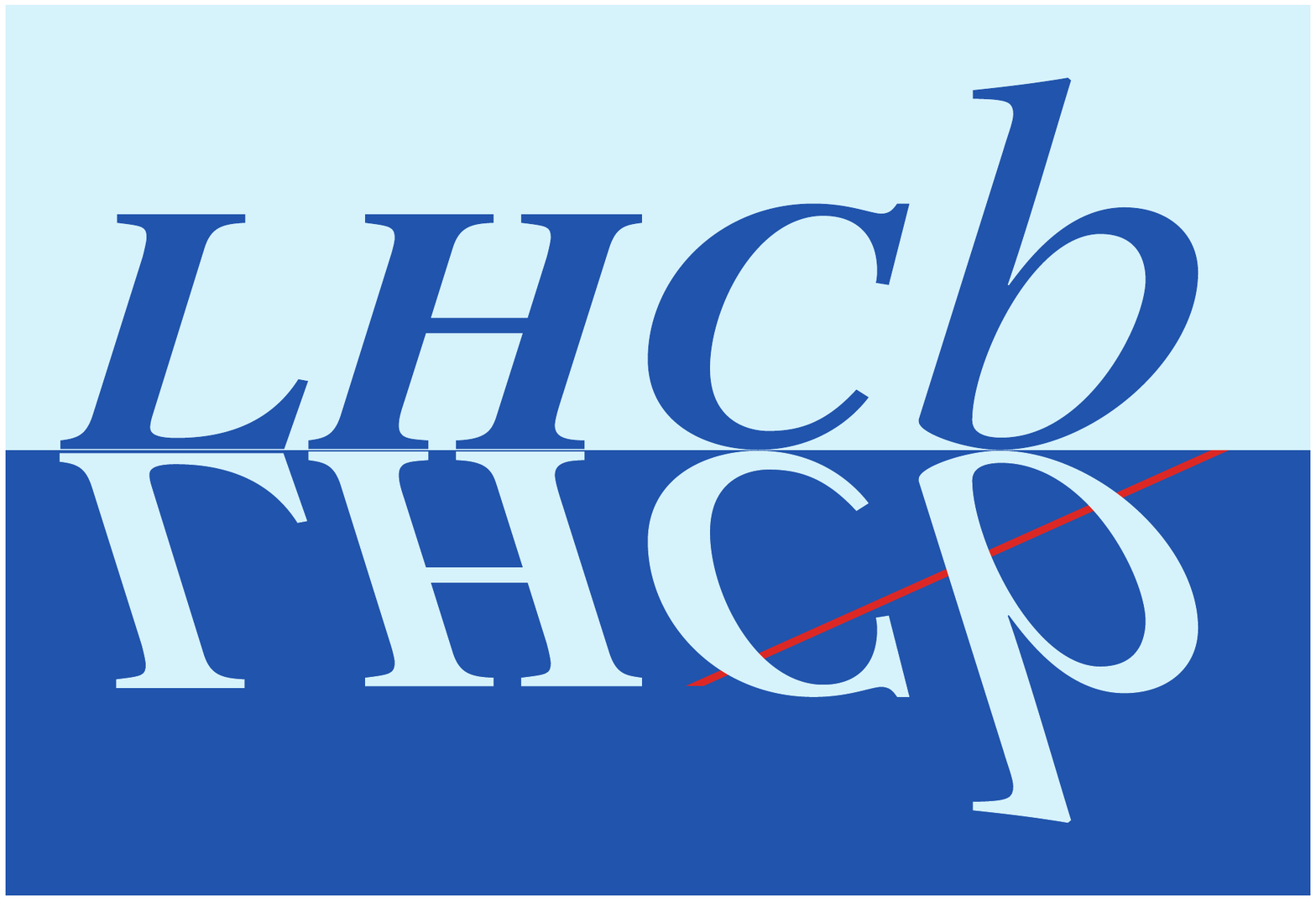}} & &}%
{\vspace*{-1.2cm}\mbox{\!\!\!\includegraphics[width=.12\textwidth]{lhcb-logo.eps}} & &}\\
 & & CERN-PH-EP-2015-202 \\
 & & LHCb-PAPER-2015-036 \\
 & & August 13, 2015 \\
 & & \\
\end{tabular*}

\vspace*{4.0cm}

{\bf\boldmath\huge
\begin{center}
  Search for hidden-sector bosons in $B^0 \!\to K^{*0}\mu^+\mu^-$ decays
\end{center}
}

\vspace*{0.1cm}

\begin{center}
The LHCb collaboration\footnote{Authors are listed at the end of this Letter.}
\end{center}

\vspace{\fill}

\begin{abstract}
  \noindent
  A search is presented for hidden-sector bosons, $\chi$, 
produced in the decay ${B^0\!\to K^*(892)^0\chi}$, with $K^*(892)^0\!\to K^{+}\pi^{-}$ and $\chi\!\to\mu^+\mu^-$.
The search is performed using $pp$-collision data corresponding to 3.0\invfb  collected with the LHCb detector.
 No significant signal is observed in the accessible mass range $214 \leq m({\chi}) \leq 4350 \mev$, and upper limits are placed on the branching fraction product
$\mathcal{B}(B^0\!\to K^*(892)^0\chi)\times\mathcal{B}(\chi\!\to\mu^+\mu^-)$ 
as a function of the mass and lifetime of the $\chi$ boson.
These limits are of the order of $10^{-9}$ for $\chi$ lifetimes less than 100\,ps over most of the \mx range, and place the most stringent constraints to date on 
many theories 
that predict the existence of additional low-mass bosons.
\end{abstract}

\vspace*{0.5cm}

\begin{center}
  Published as Physical Review Letters {\bf 115}  (2015) 161802.
\end{center}

\vspace{\fill}

{\footnotesize
\centerline{\copyright~CERN on behalf of the \lhcb collaboration, license \href{http://creativecommons.org/licenses/by/4.0/}{CC-BY-4.0}.}}
\vspace*{2mm}

\end{titlepage}

\newpage
\setcounter{page}{2}
\mbox{~}
\newpage

\renewcommand{\thefootnote}{\arabic{footnote}}
\setcounter{footnote}{0}

\pagestyle{plain}
\setcounter{page}{1}
\pagenumbering{arabic}


\clearpage

Interest has been rekindled in hidden-sector theories~\cite{Essig:2013lka}, 
motivated by the current lack of evidence for a dark matter particle candidate 
and by various cosmic-ray anomalies~\cite{Weidenspointner:2006nua,Chang:2008aa,Adriani:2008zr,2011PhRvL.106t1101A,Adriani:2013uda,FermiLAT:2011ab,PhysRevLett.113.121102}. 
These theories postulate that dark matter particles interact feebly with all known particles, which is why they have escaped detection.
Such interactions can be generated in theories where hidden-sector particles are singlet states under the Standard Model (SM) gauge interactions.
Coupling between the SM and hidden-sector particles may then arise via mixing between the hidden-sector field and any SM field with an associated particle that is not charged under the electromagnetic or strong interaction  (the Higgs and $Z$ bosons, the photon, and the neutrinos).  
This mixing could provide a so-called portal through which a hidden-sector particle, $\chi$, may be produced if kinematically allowed.

Many theories predict that \tev-scale dark matter particles interact via \gev-scale bosons~\cite{ArkaniHamed:2008qn,Pospelov:2008jd,Cheung:2009qd}  ($c=1$ throughout this Letter). 
Previous searches for such \gev-scale particles have been performed using large data samples from 
many types of experiments (see Ref.~\cite{Alekhin:2015byh} for a summary). 
These searches have placed stringent constraints on the properties of the hidden-sector photon and neutrino portals; however, the constraints on the axial-vector and scalar portals are significantly weaker.

One class of models involving the scalar portal hypothesizes that 
such a $\chi$ field was responsible for an inflationary period in the early universe~\cite{Bezrukov:2009yw}, and may have generated the baryon asymmetry observed today~\cite{Hertzberg:2013mba,Hertzberg:2013jba}.
The associated inflaton particle is expected to have a mass in the range $270\lesssim\mx\lesssim1800\mev$~\cite{Bezrukov:2009yw}.
Another class of models invokes 
the axial-vector portal in theories of dark matter that seek to address
the cosmic-ray anomalies, 
and to explain the suppression of charge-parity ($C\!P$) violation in strong interactions~\cite{Peccei:2006as}. 
These theories postulate an additional fundamental symmetry,  
the spontaneous breaking of which results in a particle called the axion~\cite{PhysRevLett.38.1440}.
To couple the axion portal to a hidden sector containing a \tev-scale dark matter
particle, while also explaining the suppression of $C\!P$ violation in strong interactions, Ref.~\cite{Nomura:2008ru} proposes an axion with $360\lesssim\mx\lesssim800\mev$ and an energy scale, $f(\chi)$, at which the symmetry is broken in the range $1\lesssim f({\chi}) \lesssim 3\tev$.  A broader range of \mx and $f(\chi)$ values is allowed in other dark matter scenarios involving axion(-like) states~\cite{Mardon:2009gw,Freytsis:2009ct,Hooper:2009gm}.

\begin{figure}
  \begin{center}
    \includegraphics[scale=1]{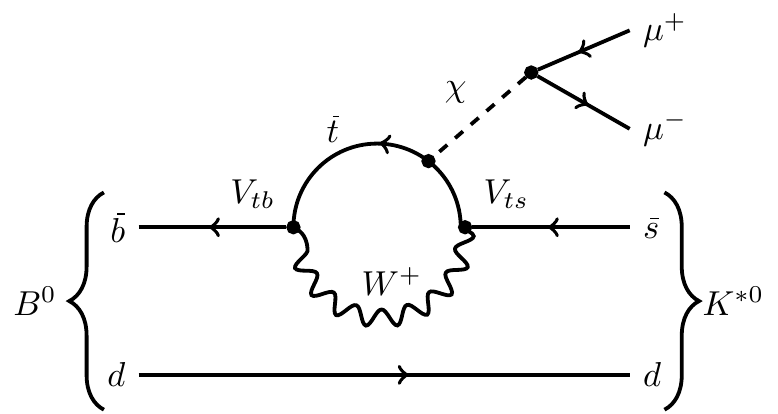}
    \caption{
      Feynman diagram for the decay \sigdecay,  with $\chi \! \to \mu^+\mu^-$.
      \label{fig1}}
  \end{center}
\end{figure}

This Letter reports a search for a hidden-sector boson produced in the decay $B^0\!\to K^{*0}\chi$, with $\chi \! \to \mu^+\mu^-$ and $K^{*0} \!\to K^+\pi^-$ (throughout this Letter, $K^{*0} \equiv K^{*}(892)^0$ and the inclusion of charge-conjugate processes is implied).
Enhanced sensitivity to hidden-sector bosons 
arises because the $b\!\to s$ transition is mediated by a top quark loop at leading order (see Fig.~\ref{fig1}).
Therefore, a $\chi$ boson with $2m(\mu) < \mx < m(B^0)-m(K^{*0})$ and a sizable top quark coupling, {\em e.g.} obtained via mixing with the Higgs sector, could be produced at a substantial rate in such decays.  
The \sigdecay decay is chosen instead of $B^+\!\to K^+\chi$, since better $\chi$ decay time resolution is obtained due to the presence of the $K^+\pi^-$ vertex, and because there is less background contamination.  
The data used correspond to integrated luminosities of 1.0 and $2.0\fb^{-1}$ collected at center-of-mass energies of $\sqrt{s}=7$ and 8\tev in $pp$ collisions with the LHCb detector.
This is the first dedicated search over a large mass range for a hidden-sector boson in a decay mediated by a $b\!\to s$ transition at leading order, and the most sensitive search to date over the entire accessible mass range.  
Previous limits set on $\chi$ boson production in such decays have either focused on a limited mass range~\cite{Hyun:2010an}, or have been obtained from more general searches for long-lived particles~\cite{Lees:2015rxq}.

The \lhcb detector is a single-arm forward
spectrometer covering the \mbox{pseudorapidity} range $2<\eta <5$,
designed for the study of particles containing \bquark or \cquark
quarks~\cite{Alves:2008zz,Aaij:2014jba}.
The detector includes a high-precision charged-particle tracking system for measuring momenta~\cite{LHCbVELOGroup:2014uea,LHCb-DP-2013-003}; 
two ring-imaging Cherenkov detectors for distinguishing charged hadrons~\cite{LHCb-DP-2012-003};
a calorimeter system for identifying photons, electrons, and hadrons;
and a system for identifying muons~\cite{LHCb-DP-2012-002}.
The trigger consists of a
hardware stage, based on information from the calorimeter and muon systems, 
followed by a software stage, which applies a full event
reconstruction~\cite{LHCb-DP-2012-004}. 
The selection of \sigdecay candidates in the software trigger requires the presence of
a vertex identified by a multivariate algorithm~\cite{BBDT} as being consistent with the decay of a \bquark hadron.
Alternatively, candidates may be selected based on the presence of a displaced dimuon vertex, or the presence of a muon with large transverse momentum (\pt) and large impact parameter (IP), defined as the minimum track distance with respect to any $pp$-interaction vertex (PV).  
Only tracks with segments reconstructed in the first charged-particle detector, which surrounds the interaction region and is about 1\,m in length~\cite{LHCbVELOGroup:2014uea}, can satisfy these trigger requirements; 
therefore, the $\chi$ boson is required to decay well within this detector. 
In the simulation, $pp$ collisions are generated following Refs.~\cite{Sjostrand:2007gs,LHCb-PROC-2010-056,Lange:2001uf,Golonka:2005pn}, and the interactions of the outgoing particles with the detector are modelled as in Refs.~\cite{Allison:2006ve, *Agostinelli:2002hh,LHCb-PROC-2011-006}. 

A search is conducted, following Ref.~\cite{Mike}, by scanning the \mmm distribution for an excess of $\chi$ signal candidates over the expected background.  
In order to avoid experimenter bias, all aspects of the search are fixed without examining those \sigdecay candidates which have an invariant mass consistent with the known $B^0$ mass~\cite{PDG2014}.  
The step sizes in \mx are 
$\sigma[\mmm]/2$, where $\sigma[\mmm]$ is the dimuon mass resolution.
Signal candidates satisfy  $|\mmm-\mx| < 2\sigma[\mmm]$, while the background is estimated by interpolating the yields in the sidebands starting at $3\sigma[\mmm]$ from \mx.  
With $m(K^+\pi^-\mu^+\mu^-)$ constrained~\cite{Hulsbergen:2005pu} to the known $B^0$ mass, $\sigma[\mmm]$ is less than 8\mev over the entire \mmm range, and is as small as 2\mev below 220\mev. 
The statistical test at each \mx is based on the profile likelihood ratio of Poisson-process hypotheses with and without a signal contribution~\cite{Cowan:2010js}.  
The uncertainty on the background interpolation is modeled by a Gaussian term in the likelihood (see Ref.~\cite{Mike} for details).

The $\chi\!\to\mu^+\mu^-$ decay vertex is permitted, but not required, to be displaced from the \sigdecay decay vertex.  Two 
regions of reconstructed dimuon lifetime, \tmm, are defined for each \mx considered in the search: a prompt region, $|\tmm| < 3\sigma[\tmm]$, and a displaced region, $\tmm > 3\sigma[\tmm]$. 
The lifetime resolution is about $0.2\ps$ for $\mmm \gtrsim 250\mev$, 
and $1\ps$ near $2m(\mu)$.  
 The joint likelihood is formed from the product of the likelihoods for candidates populating the prompt and displaced regions, since   
no assumption is made about \tx.
Narrow resonances are vetoed by excluding the regions near the $\omega$, $\phi$, $J/\psi$, $\psi(2S)$ and $\psi(3770)$ resonances.  These regions are removed in both the prompt and displaced samples to avoid contamination from unassociated dimuon and $K^{*0}$ resonances.

The branching fraction product $\mathcal{B}(B^0\!\to K^{*0}\chi(\mu^+\mu^-))\equiv\mathcal{B}(\sigdecay)\times\mathcal{B}(\chi\!\to\mu^+\mu^-)$ is measured relative to $\mathcal{B}(\smdecay)$, where the normalization sample is taken from the prompt region and restricted to $1.1 < \mmmsq < 6.0\gev^2$.  
This normalization decay is chosen since the detector response is similar to that for the \sigdecay decay, and because the hidden-sector theory parameters can be obtained from the ratio $\mathcal{B}(B^0\!\to K^{*0}\chi(\mu^+\mu^-))/\mathcal{B}(\smdecay)$ with reduced theoretical uncertainty.  
Correlations between the yields of a possible signal in the prompt $1.1 < \mmmsq < 6.0\gev^2$ region and the normalization decay are at most a few percent and are ignored.

The selection is similar to that of Ref.~\cite{LHCb-PAPER-2013-019} with the exception that the $K^{*0}$ and dimuon candidates are not required to share a common vertex.
Signal candidates are required to satisfy a set of loose requirements:
the $B^0$, $K^{*0}$ and $\chi$ decay vertices must all be separated from any PV and be of good quality;
the $B^0$ IP must be small, while the IP of the kaon, pion and muons must be large;
the angle between the $B^0$ momentum vector and the vector between the associated PV and the $B^0$ decay vertex must be small;
and
the kaon, pion and muons must each satisfy loose particle identification requirements.
Candidates are retained if $m(K^+\pi^-)$ is within 100\mev of the known $K^{*0}$ mass~\cite{PDG2014}.

A multivariate selection is applied to reduce the background further.
The uBoost algorithm~\cite{Stevens:2013dya} is employed
to ensure that the performance is nearly independent of \mx and \tx. 
The inputs to the algorithm include $\pt(B^0)$, various topological features of the decay, the muon identification quality, and an isolation criterion~\cite{Aaij:2029609} designed to suppress backgrounds from  partially reconstructed decays.
Data from the high-mass sideband, $150 < m(K^+\pi^-\mu^+\mu^-) - m(B^0) < 500\mev$, are used to represent the background in the training, while simulated samples generated with \mx values of 214, 1000, and 4000\mev, and \tx large enough to populate the full reconstructible region, are used for the signal. 
The multivariate selection requirement is determined by maximizing
the figure of merit of Ref.~\cite{Punzi:2003bu} for finding a signal with a significance of five standard deviations.  
This results in a signal selection efficiency of 85\% with a background rejection of 92\% on average. 
The uBoost algorithm is validated using ten additional signal samples generated with various other \mx and \tx values.  
The performance is consistent for all samples.

Peaking backgrounds that survive the multivariate selection are vetoed explicitly.
A small number of $B_s^0 \!\to \phi(K^+K^-)\mu^+\mu^-$ decays are removed by rejecting $K^+\pi^-$ candidates that are consistent with the decay $\phi\!\to K^+K^-$ if the $\pi^-$ is assumed to be a misidentified $K^-$.  
A similar veto is applied that removes about 250  $\PLambda_b^0 \!\to pK^-\mu^+\mu^-$ decays.
Candidates are also rejected if the dimuon system is  consistent with any of the following decays:
$K_S^0 \!\to \pi^+\pi^-$, where the pions decay in flight to muons;
$\PLambda^0 \!\to p\pi^-$, where the pion decays in flight and the proton is misidentified as a muon;
and $\Dzb \! \to K^+\pi^-$, where the kaon and pion decay in flight.
All other particle-misidentification backgrounds are negligible.  

\begin{figure}
  \begin{center}
    \includegraphics[width=0.48\textwidth]{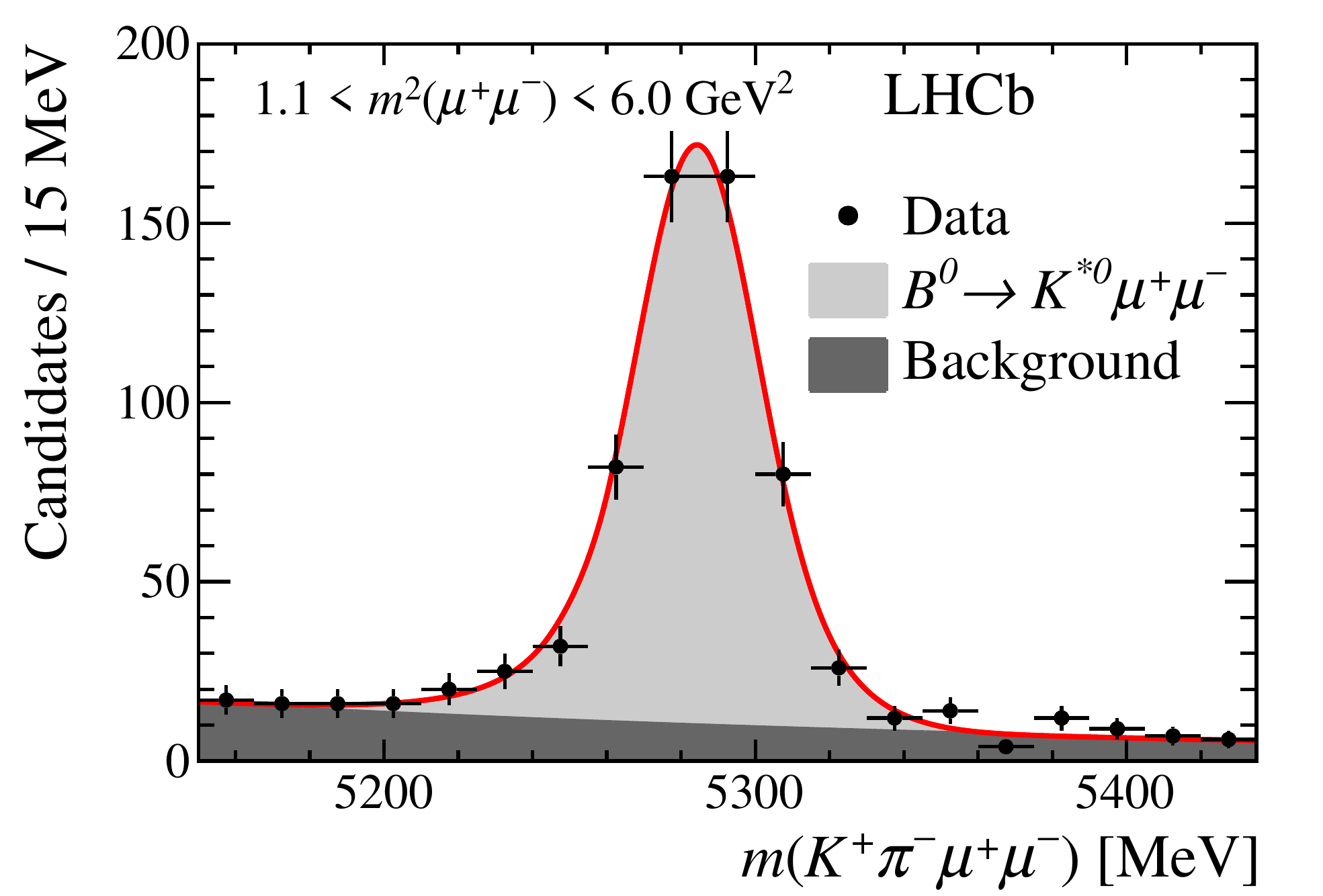}
    \caption{
      Invariant mass spectrum with fit overlaid for all prompt \smdecay candidates with $1.1 < \mmmsq < 6.0\gev^2$.
      \label{fig2}}
  \end{center}
\end{figure}

Figure~\ref{fig2} shows the $K^+\pi^-\mu^+\mu^-$ mass distribution for all prompt candidates that satisfy the full selection in the region $1.1 < \mmmsq < 6.0\gev^2$.  An unbinned extended maximum likelihood fit is performed to obtain the \smdecay yield.  The signal model is obtained from data using the subset of prompt candidates with \mmm in the \jpsi region, 
where the background is $\mathcal{O}(10^{-3})$.  A small correction, obtained from simulation, is applied to account for the difference in signal shape expected in the ${1.1 < \mmmsq < 6.0\gev^2}$ region.  The background model is an exponential function.  Several alternative background models are considered, with the largest shift observed   in the signal yield (1\%) assigned as a systematic uncertainty.  The $S$-wave fraction ({\em i.e.}\ not a $K^{*0}$ meson) of the $K\pi$ system within the selected $K\pi$ mass range is $(4\pm4)\%$~\cite{LHCb-PAPER-2013-019}.  The yield of the normalization mode is  $N(\smdecay)= 506\pm33$, where the uncertainty includes both statistical and systematic contributions.

Probability density functions, obtained from the data using splines, are used to generate simulated data sets under the no-signal hypothesis from which the global significance of any $\chi$ signal is obtained~\cite{Mike}.  
For this the data are collected in the prompt region 
into wide bins with a width of 200\mev, and into a total of three bins in the displaced region.  Simulated events show that the presence of a narrow $\chi$ signal anywhere in the \mx-\tx plane, whose local significance is $5\sigma$, would not produce a significant excess in these wide-binned data.

\begin{figure}
  \begin{center}
    \includegraphics[width=1.0\textwidth]{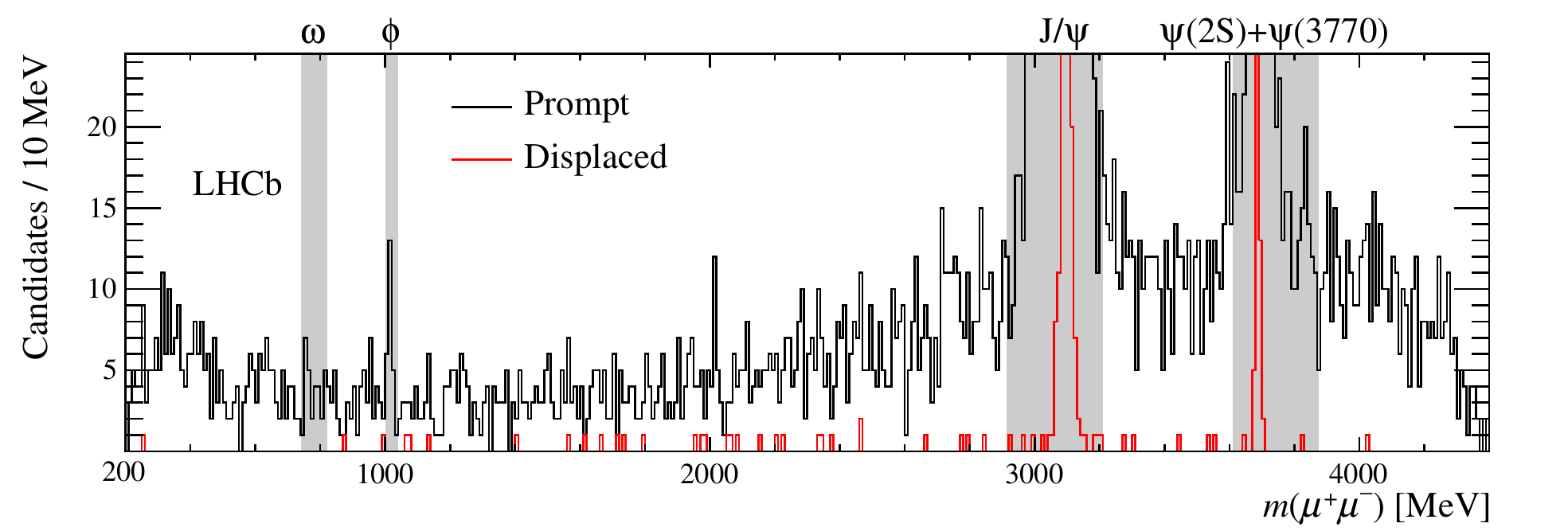}
    \caption{
      Distribution of \mmm in the (black) prompt and (red) displaced regions.  The shaded bands denote regions where no search is performed due to (possible) resonance contributions.  The \jpsi, $\psi(2S)$ and $\psi(3770)$ peaks are suppressed to better display the search region.
      \label{fig3}}
  \end{center}
\end{figure}

Figure~\ref{fig3} shows the \mmm distributions in both the prompt and displaced regions for candidates whose invariant mass is within $50\mev$ of the known $B^0$ mass.  
The most significant local excess occurs for $\mx=253\mev$,  where in the prompt region 11\,(6.2) candidates are observed (expected), while the displaced region contains a single candidate which is the only displaced candidate below $m(\omega)$.  
The $p$-value of the no-signal hypothesis is about 80\%, showing that no evidence is found for a hidden-sector boson.

To set upper limits on $\mathcal{B}(\sigdecay(\mu^+\mu^-))$, various sources of systematic uncertainty are considered. 
The limits are set using the profile likelihood technique~\cite{Rolke:2004mj}, in which systematic uncertainties are handled by including additional Gaussian terms in the likelihood~\cite{Mike}.  
Since no contamination from the $\omega$ or $\phi$ resonance is found in the displaced region, upper limits are set in these \mx regions for $\tx > 1\ps$.

Many uncertainties cancel to a good approximation because the signal and normalization decays share the same final state. 
The dominant uncertainty on the efficiency ratio $\epsilon(\sigdecay(\mu^+\mu^-))/\epsilon(\smdecay)$, which is taken from simulation, arises due to its dependence on \tmm.
The simulation is validated by comparing $\tau(\pi^+\pi^-)$ distributions between $B^0 \!\to \jpsi K_S^0(\pi^+\pi^-)$ decays reconstructed in simulated and experimental data in bins of $K_S^0$ momentum.
The distributions in data and simulation are consistent in each bin, 
and the per-bin statistical precision (5\%) is assigned as systematic uncertainty. 

The uncertainty on the efficiency for a signal candidate to be reconstructed within a given \mmm signal window, due to mismodeling of $\sigma[\mmm]$, is determined to be 1\%
based on a comparison of the \jpsi peak between $B^0 \!\to \jpsi(\mu^+\mu^-) K^{*0}$ decays in simulated and experimental data. 
A similar comparison for $\sigma[\tmm]$ shows that the uncertainty on 
the fraction of signal candidates expected to be reconstructed in the prompt and displaced regions 
is negligible.
Finally, the efficiency for the normalization mode is determined 
using the measured angular distribution~\cite{LHCb-CONF-2015-002}, which is varied within the uncertainties yielding an uncertainty in the normalization-mode efficiency of 1\%. 
The individual contributions are summed in quadrature giving a total systematic uncertainty of 8\%.

The spin of the hidden-sector boson determines the angular distribution of the decay and, therefore, affects the efficiency.
The upper limits are set assuming spin zero.
For a spin-one $\chi$ boson produced unpolarized in the decay, the sensitivity is about 10--20\% better than for the spin-zero case.  
The dependence on the polarization in the spin-one case is provided as supplemental material to this Letter~\cite{supp}. 

\begin{figure}
  \begin{center}
   \includegraphics[width=1.0\textwidth]{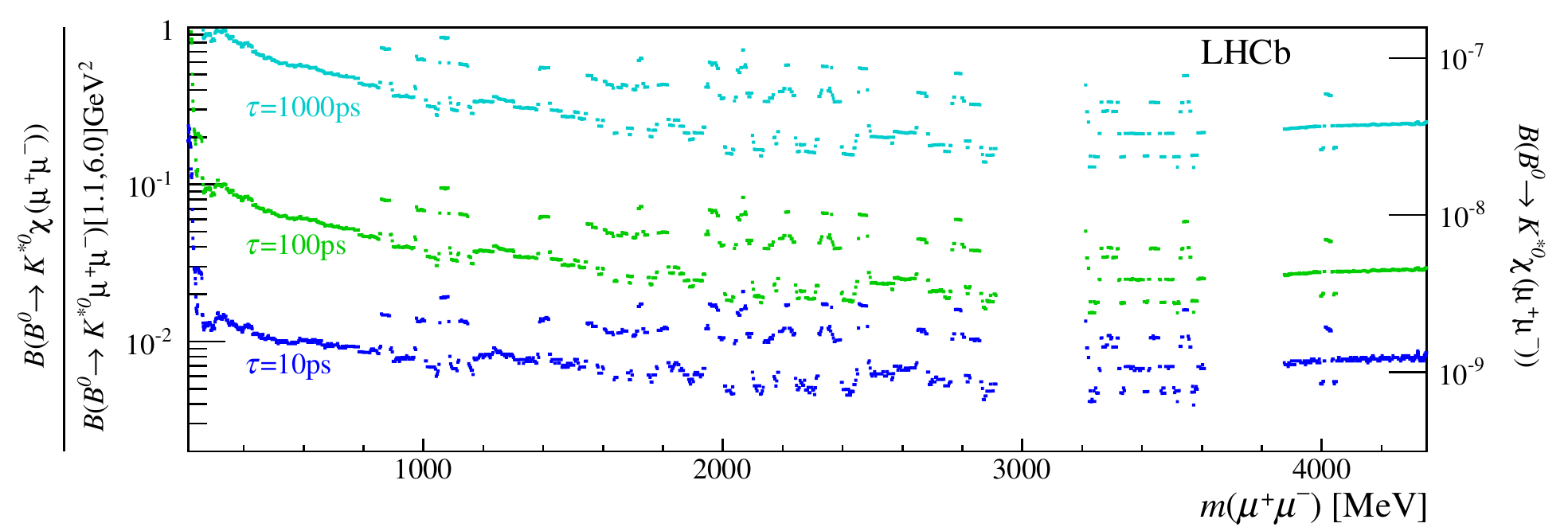}
  \caption{
      Upper limits at 95\% CL for (left axis)
      $\mathcal{B}(\sigdecay(\mu^+\mu^-))/\mathcal{B}(\smdecay)$, with
      \smdecay in $1.1 < \mmmsq < 6.0\gev^2$, and (right axis)
      $\mathcal{B}(\sigdecay(\mu^+\mu^-))$.
	The sparseness of the data leads to rapid fluctuations in the limits.  
	Excluding the region near $2m(\mu)$, 
        the relative limits for $\tau < 10\ps$ are between 0.005--0.05 and 
        all relative limits for $\tau \leq 1000\ps$ are less than one. 
      \label{fig4}}
  \end{center}
\end{figure}

Figure~\ref{fig4} shows the upper limits on $\mathcal{B}(\sigdecay(\mu^+\mu^-))$, relative to $\mathcal{B}(\smdecay)$ in the $1.1 < \mmmsq < 6.0\gev^2$ region, set at the 95\% confidence level (CL) for several values of \tx; limits as functions of \tx are provided as supplemental material to this Letter.
The limits become less stringent for $\tx \gtrsim 10\ps$, as the probability of the $\chi$ boson decaying within the first charged-particle detector decreases.
The branching fraction $\mathcal{B}(\smdecay)=(1.6\pm0.3)\times10^{-7}$~\cite{LHCb-PAPER-2013-019} is used to obtain upper limits on $\mathcal{B}(\sigdecay(\mu^+\mu^-))$, which are also shown in  Fig.~\ref{fig4}. 
Due to the uncertainty on the normalization-mode branching fraction, there is not a one-to-one mapping  between the two axes in the figure; however, the absolute limits shown are accurate to about 2\%.

\begin{figure}
  \begin{center}
    \includegraphics[width=0.48\textwidth]{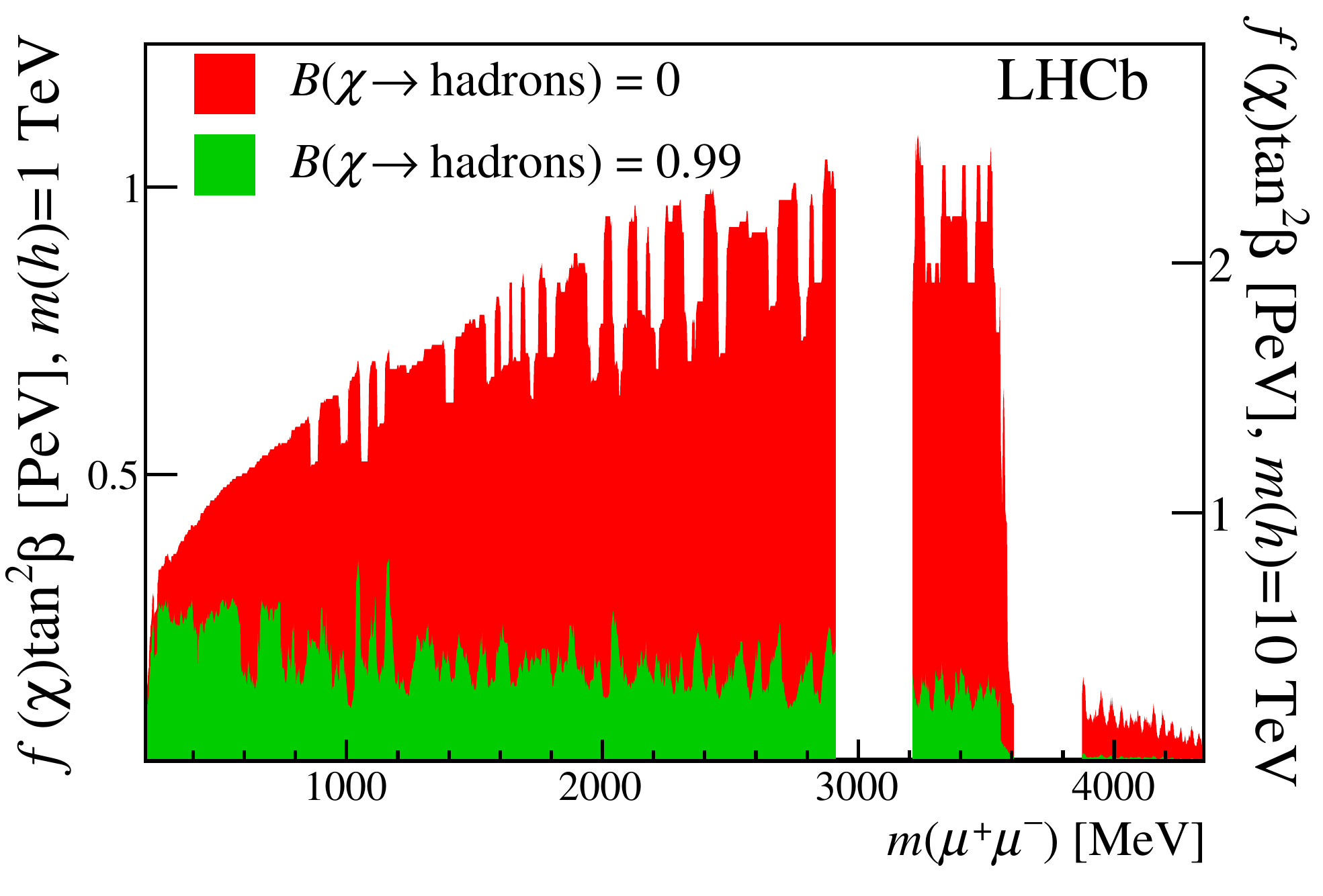}
    \includegraphics[width=0.48\textwidth]{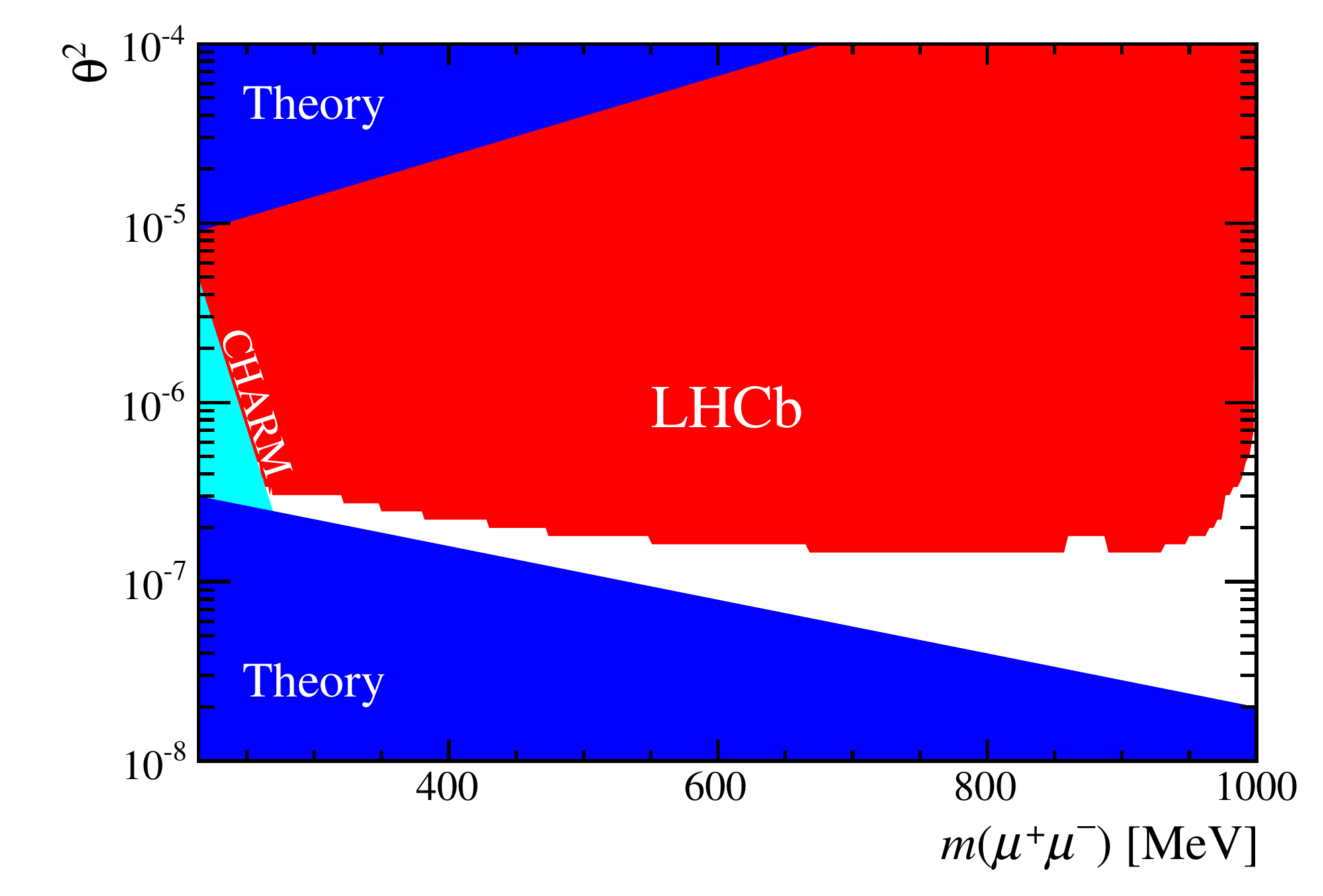}
    \caption{
      Exclusion regions at 95\% CL: (left) constraints on the axion model of Ref.~\cite{Freytsis:2009ct};
      (right) constraints on the inflaton model of Ref.~\cite{Bezrukov:2014nza}. 
      The regions excluded by the theory~~\cite{Bezrukov:2014nza} and by the CHARM experiment~\cite{Bergsma:1985qz} are also shown.
      \label{fig5}}
  \end{center}
\end{figure}

Figure~\ref{fig5} shows exclusion regions for the DFSZ~\cite{Dine:1981rt,Zhitnitsky:1980tq} axion model of Ref.~\cite{Freytsis:2009ct} set in the limit of large ratio of Higgs-doublet vacuum expectation values, $\tan{\beta} \gtrsim 3$, for charged-Higgs masses $m(h)=1$ and 10\tev (this choice of restricted parameter space is made for ease of graphical presentation). 
The constraints scale as $\log{(m(h)/\tev)}$ for $m(h) \gtrsim 800\gev$.   
The branching fraction of the axion into hadrons varies greatly in different models.
Figure~\ref{fig5} shows the results for two extreme cases: $\mathcal{B}(\chi \!\to {\rm hadrons}) = 0$ and 0.99.
While $\mathcal{B}(\chi \!\to \mu^+\mu^-)$ is 100 times larger when $\mathcal{B}(\chi \!\to {\rm hadrons}) = 0$,
\tx is also larger, which 
results in the model probing the region where the upper limits are weaker.
The constraints are loose for $\mx > 2m(\tau)$, since the axion preferentially decays into $\tau^+\tau^-$ if kinematically allowed; otherwise the exclusions reach the PeV scale.

Figure~\ref{fig5} also shows exclusion regions for the inflaton model of Ref.~\cite{Bezrukov:2014nza}, which only considers  $\mx < 1\gev$.  The branching fraction into hadrons
is taken directly from Ref.~\cite{Bezrukov:2014nza} and, as in the axion model, is highly uncertain but this does not greatly affect the sensitivity of this search.  Constraints are placed on the mixing angle between the Higgs and inflaton fields, $\theta$, 
which 
exclude most of the previously allowed region.

In summary, 
no evidence for a signal is observed, and upper limits are placed on 
$\mathcal{B}(\sigdecay)\times\mathcal{B}(\chi\!\to\mu^+\mu^-)$.  
This is the first dedicated search over a large mass range for a hidden-sector boson in a decay mediated by a $b\!\to s$ transition at leading order, and the most sensitive search to date over the entire accessible mass range.  
Stringent constraints are placed on theories that predict the existence of additional scalar or axial-vector fields.

\section*{Acknowledgments}

\noindent
We express our gratitude to our colleagues in the CERN
accelerator departments for the excellent performance of the LHC. We
thank the technical and administrative staff at the LHCb
institutes. We acknowledge support from CERN and from the national
agencies: CAPES, CNPq, FAPERJ and FINEP (Brazil); NSFC (China);
CNRS/IN2P3 (France); BMBF, DFG, HGF and MPG (Germany); INFN (Italy);
FOM and NWO (The Netherlands); MNiSW and NCN (Poland); MEN/IFA (Romania);
MinES and FANO (Russia); MinECo (Spain); SNSF and SER (Switzerland);
NASU (Ukraine); STFC (United Kingdom); NSF (USA).
The Tier1 computing centres are supported by IN2P3 (France), KIT and BMBF
(Germany), INFN (Italy), NWO and SURF (The Netherlands), PIC (Spain), GridPP
(United Kingdom).
We are indebted to the communities behind the multiple open
source software packages on which we depend. We are also thankful for the
computing resources and the access to software R\&D tools provided by Yandex LLC (Russia).
Individual groups or members have received support from
EPLANET, Marie Sk\l{}odowska-Curie Actions and ERC (European Union),
Conseil g\'{e}n\'{e}ral de Haute-Savoie, Labex ENIGMASS and OCEVU,
R\'{e}gion Auvergne (France), RFBR (Russia), XuntaGal and GENCAT (Spain), Royal Society and Royal
Commission for the Exhibition of 1851 (United Kingdom).

\addcontentsline{toc}{section}{References}
\setboolean{inbibliography}{true}
\bibliographystyle{LHCb}
\bibliography{paper.bbl}

\clearpage

\section*{Supplemental Material}

\begin{figure}[h!]
  \begin{center}
    \includegraphics[width=0.49\textwidth]{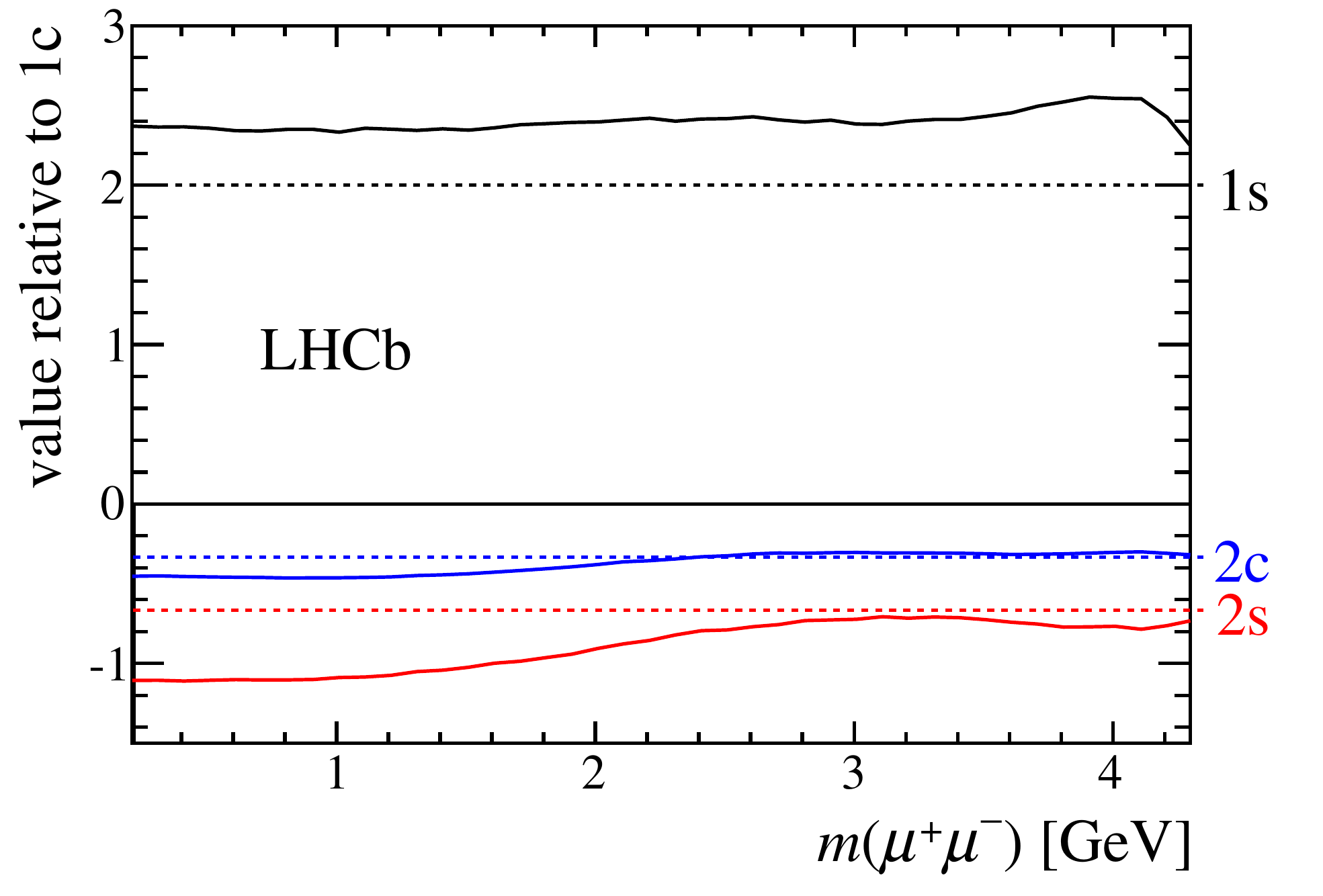}
    \includegraphics[width=0.49\textwidth]{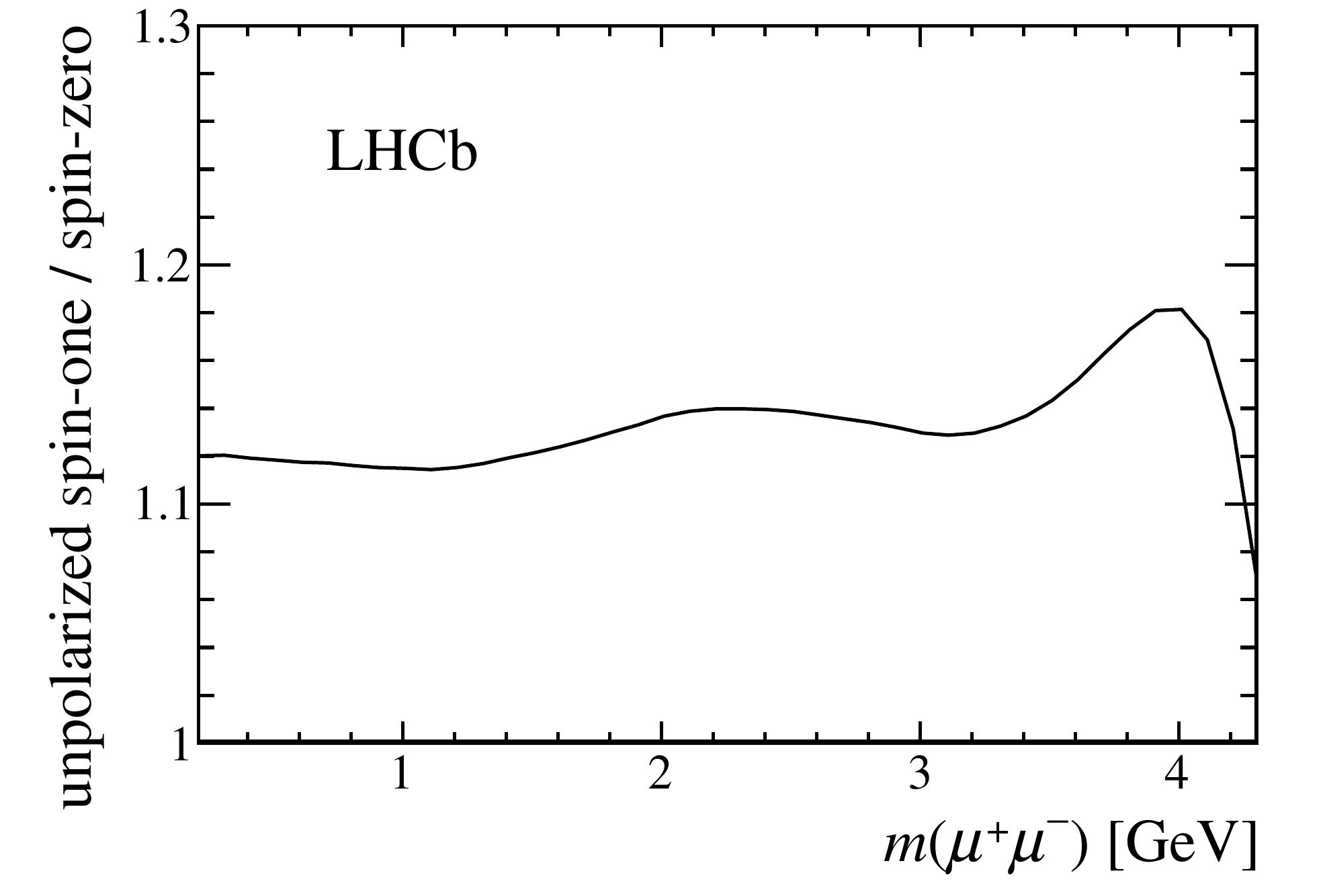}
    \caption{
     \label{fig1S}
      (left) Integral values for 1s, 2s and 2c relative to the value for 1c (see text for details).  The dashed lines show the values in the absence of inefficiency.  (right) Ratio of the efficiency for an unpolarized spin-one boson to that of a spin-zero boson.  
      }
  \end{center}
\end{figure}

The limits reported in the Letter assume a spin-zero hidden-sector boson.  To convert these into limits for a spin-one boson, the ratio of efficiencies for the spin-one to spin-zero cases must be accounted for.   Determining this ratio involves integrals of the form
\begin{equation}
\frac{\int f_j(\vec{\Omega}) \epsilon(\vec{\Omega}, m^2(\mu^+\mu^-)) {\rm d} \vec{\Omega}}{\int f_{1c}(\vec{\Omega}) \epsilon(\vec{\Omega}, m^2(\mu^+\mu^-)) {\rm d} \vec{\Omega}}, \nonumber
\end{equation}
where $\vec{\Omega} = (\theta_K, \theta_{\ell}, \phi)$ (see Appendix~A of Ref.~[40] in the Letter for details on the angular basis), 
$\epsilon(\vec{\Omega}, m^2(\mu^+\mu^-))$ is the efficiency, $f_j(\vec{\Omega})$ are functions of the angles, 
and $f_{1c}(\vec{\Omega}) = \cos^2{\theta_K}$.  
Figure~\ref{fig1S} shows the values for
\begin{eqnarray}
f_{1s}(\vec{\Omega}) &=& \sin^2{\theta_K}, \nonumber \\
f_{2s}(\vec{\Omega}) &=& \sin^2{\theta_K}\cos{2\theta_{\ell}}, \nonumber \\
f_{2c}(\vec{\Omega}) &=& \cos^2{\theta_K}\cos{2\theta_{\ell}}. \nonumber 
\end{eqnarray}
All other integrals, each of which has a value of zero in the absence of inefficiency, have values $\mathcal{O}(0.01)$.  
Therefore, the following terms in the general angular distribution can be ignored when determining the limits:
\begin{eqnarray}
f_3(\vec{\Omega}) &=& \sin^2{\theta_K}\sin^2{\theta_{\ell}}\cos{2\phi}, \nonumber \\
f_4(\vec{\Omega}) &=& \sin{2\theta_K}\sin{2\theta_{\ell}}\cos{\phi}, \nonumber \\
f_5(\vec{\Omega}) &=& \sin{2\theta_K}\sin{\theta_{\ell}}\cos{\phi}, \nonumber \\
f_{6s}(\vec{\Omega}) &=& \sin^2{\theta_K}\cos{\theta_{\ell}}, \nonumber \\
f_{6c}(\vec{\Omega}) &=& \cos^2{\theta_K}\cos{\theta_{\ell}}, \nonumber \\
f_7(\vec{\Omega}) &=& \sin{2\theta_K}\sin{\theta_{\ell}}\sin{\phi}, \nonumber \\
f_8(\vec{\Omega}) &=& \sin{2\theta_K}\sin{2\theta_{\ell}}\sin{\phi}, \nonumber \\
f_9(\vec{\Omega}) &=& \sin^2{\theta_K}\sin^2{\theta_{\ell}}\sin{2\phi}. \nonumber 
\end{eqnarray}
Figure~\ref{fig1S} shows an example efficiency ratio for the case of a spin-one boson produced unpolarized in the decay. Since the $j=3,4,\ldots9$ terms integrate to approximately zero, this same curve applies for any theory that predicts that the longitudinal polarization fraction of the $K^{*0}$ is $F_L = 1/3$.

\begin{figure}
  \begin{center}
 \includegraphics[width=1.0\textwidth]{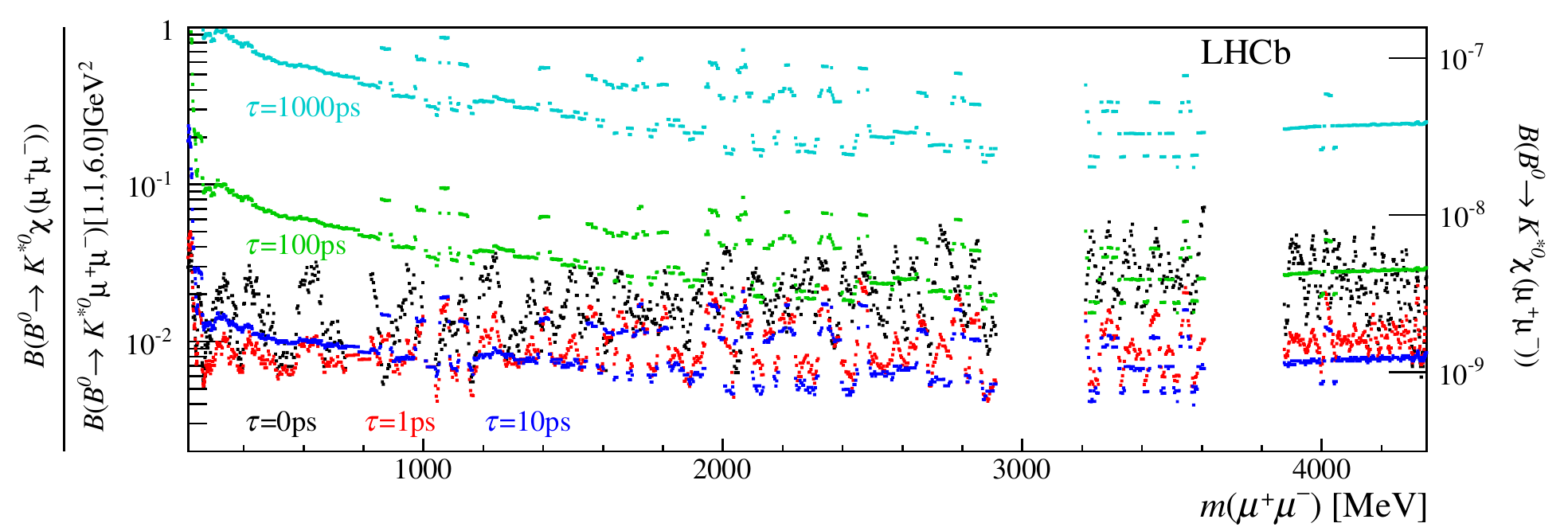}
  \caption{
      Upper limits at 95\% CL for (left axis)
      $\mathcal{B}(\sigdecay(\mu^+\mu^-))/\mathcal{B}(\smdecay)$, with
      \smdecay in $1.1 < \mmmsq < 6.0\gev^2$, and (right axis)
      $\mathcal{B}(\sigdecay(\mu^+\mu^-))$.
      Same as Fig.~4 in the Letter but including the $\tau = 0$ and 1\,ps limits.
      }
  \end{center}
\end{figure}

\begin{figure}[h!]
  \begin{center}
\includegraphics[width=1.0\textwidth]{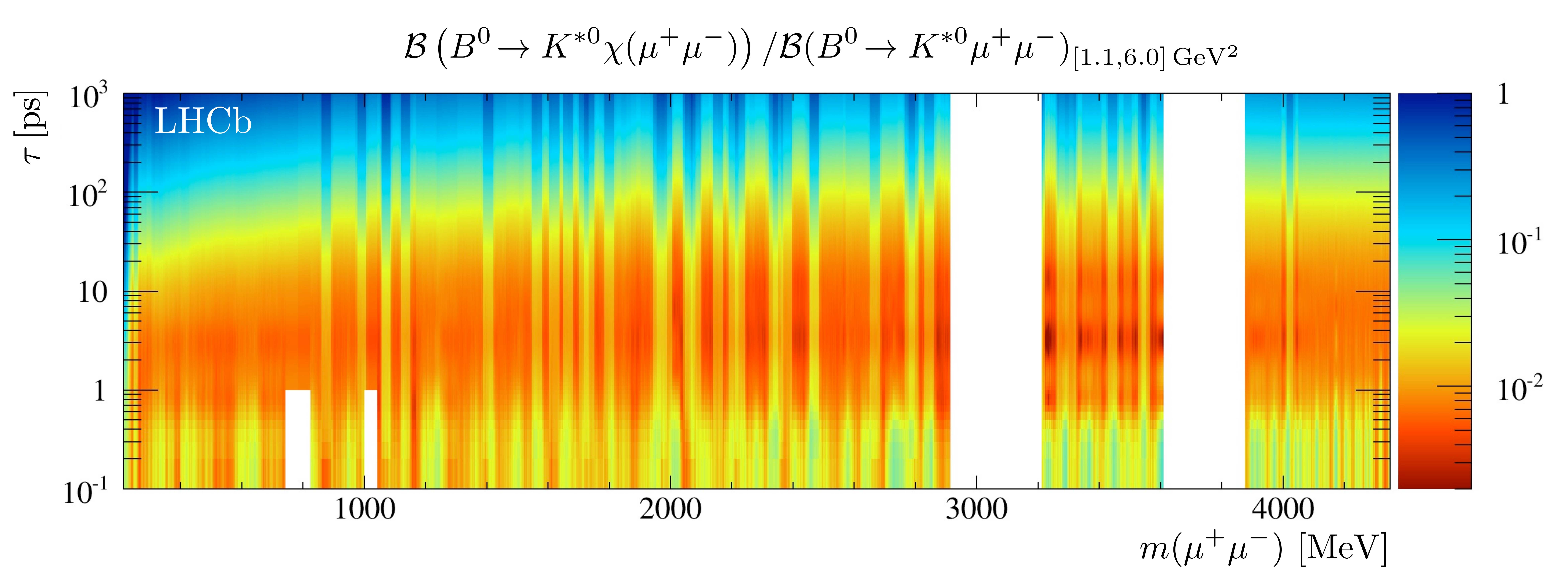}	\\
\includegraphics[width=1.0\textwidth]{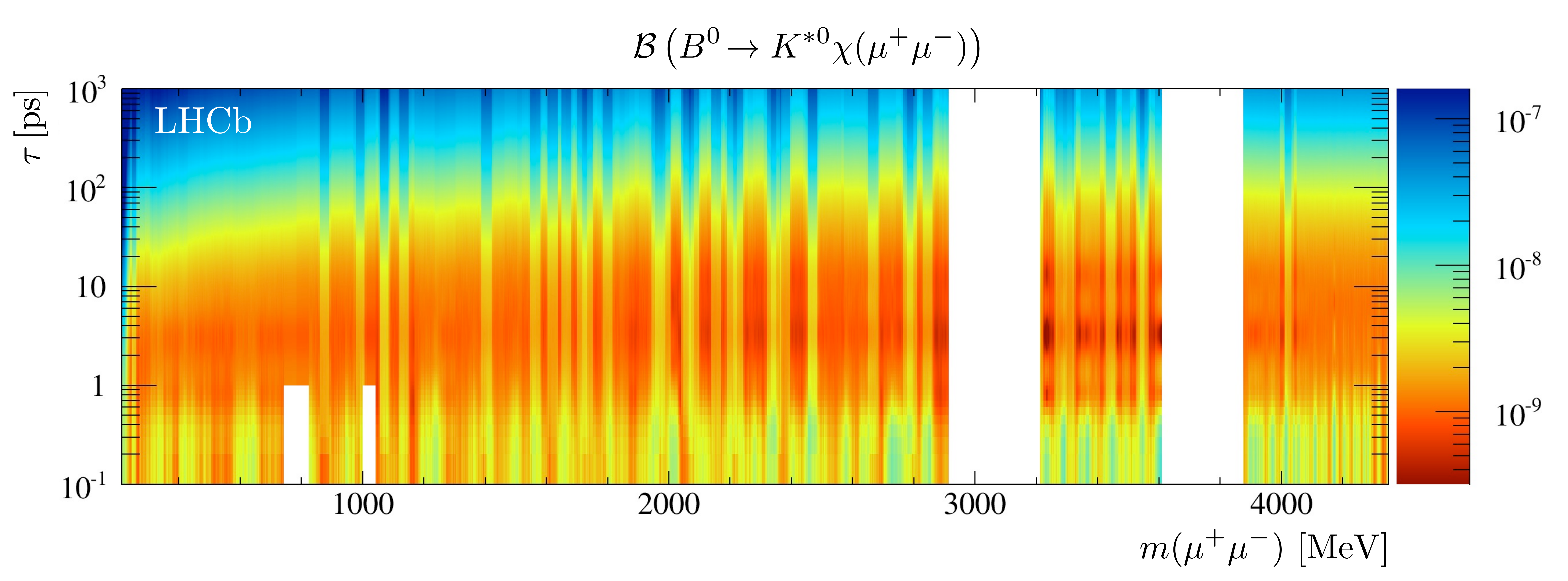}	\\
\includegraphics[width=1.0\textwidth]{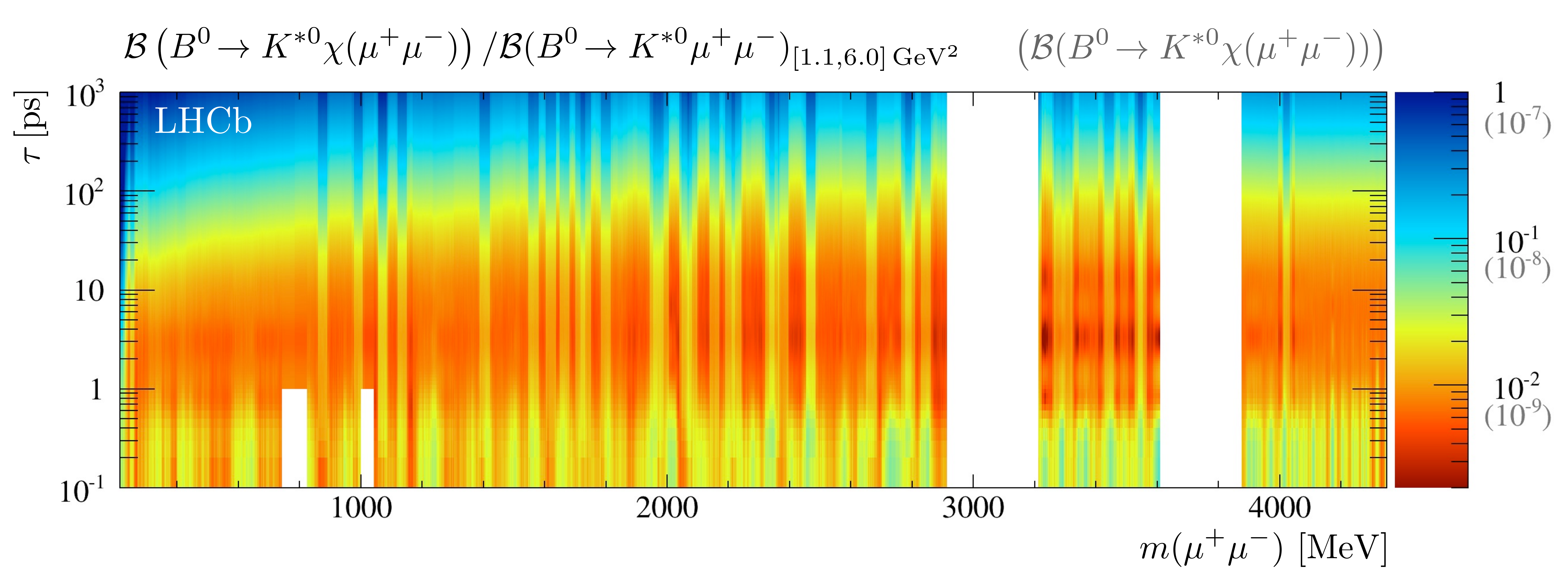}
    \caption{
      Upper limits at 95\% CL for (top) $\mathcal{B}(\sigdecay(\mu^+\mu^-))/\mathcal{B}(\smdecay)$, with \smdecay in $1.1 < \mmmsq < 6.0\gev^2$, (middle) $\mathcal{B}(\sigdecay(\mu^+\mu^-))$, and (bottom) both relative and absolute limits.
      The $\omega$ and $\phi$ resonance regions are only excluded in the prompt region.  A utility is provided to obtain these limits for any $(\mx,\tx)$ on the CERN Document Server.  
      }
  \end{center}
\end{figure}

\clearpage

\centerline{\large\bf LHCb collaboration}
\begin{flushleft}
\small
R.~Aaij$^{38}$, 
B.~Adeva$^{37}$, 
M.~Adinolfi$^{46}$, 
A.~Affolder$^{52}$, 
Z.~Ajaltouni$^{5}$, 
S.~Akar$^{6}$, 
J.~Albrecht$^{9}$, 
F.~Alessio$^{38}$, 
M.~Alexander$^{51}$, 
S.~Ali$^{41}$
% ...

{\footnotesize \it
% ...
$ ^{5}$Clermont Universit\'{e}, Universit\'{e} Blaise Pascal, CNRS/IN2P3, LPC, Clermont-Ferrand, France\\
$ ^{6}$CPPM, Aix-Marseille Universit\'{e}, CNRS/IN2P3, Marseille, France\\
% ...
$ ^{9}$Fakult\"{a}t Physik, Technische Universit\"{a}t Dortmund, Dortmund, Germany\\
% ...
$ ^{37}$Universidad de Santiago de Compostela, Santiago de Compostela, Spain\\
$ ^{38}$European Organization for Nuclear Research (CERN), Geneva, Switzerland\\
% ...
$ ^{41}$Nikhef National Institute for Subatomic Physics, Amsterdam, The Netherlands\\
% ...
$ ^{46}$H.H. Wills Physics Laboratory, University of Bristol, Bristol, United Kingdom\\
% ...
$ ^{51}$School of Physics and Astronomy, University of Glasgow, Glasgow, United Kingdom\\
$ ^{52}$Oliver Lodge Laboratory, University of Liverpool, Liverpool, United Kingdom\\
% ...
}
\end{flushleft}

\end{document}